\documentclass[12pt]{article}

\usepackage{epsfig,times} 

\author{ J.Bellandi$^{a}$, J.R.Fleitas$^{a}$ and J.Dias de Deus$^{b}$ }

\title{ Leading nucleon and inelasticity in hadron-nucleus interactions}

\date{}

\begin{document}

\maketitle

\vspace{1cm}

\begin{center}
\em{
$^{a}$ { Instituto de F\'{\i}sica Gleb Wataghin \\ Universidade Estadual de 
Campinas \\ 13083-970, Campinas SP, Brazil } \\

\vspace{0.6cm}

$^{b}$ {Departamento de F\'{\i}sica - CENTRA - IST \\  Av. Rovisco Pais, 1 - 1096 Lisboa Codex, Portugal } }

\end{center}

\vspace{3cm}

\noindent e-mail address: bellandi@ifi.unicamp.br

\vspace{0.6cm}

\noindent To appear in {\bf Journal of Physics G}.

\begin{abstract}

We present in this paper a calculation of the average proton-nucleus
inelasticity. Using an Iterative Leading Particle Model and the Glauber
model we relate the leading particle distribution in nucleon-nucleus
interactions with the respective one in nucleon-proton collisions. To
describe the leading particle distribution in nucleon-proton collisions we
use the Regge-Mueller formalism.

\end{abstract}

\newpage

\section{Introduction}

It is well known that the inelasticity is one of the most important
variables to analyse cosmic ray data on hadronic cascade and on extensive
air showers. A wide and diversified range of hadron interaction calculations
in cosmic ray and accelerator physics is strongly dependent on the
inelasticity parameter, whose energy dependence is presently in question.

Inelasticity is understood as the fraction of the available energy released
for multiple particle production in an inelastic hadron-nucleus collision.
Its value was estimated a long time ago from cosmic ray experiments \cite{1} as
being around $ 0.5$, as was later confirmed at the CERN ISR \cite{2},

The question of the energy dependence of the inelasticity was dealt by a
number of authors through different approaches , most of them in a model
dependent way. The obtained results are not consistent \cite{3},\cite{4} and the
question remains unsolved. Considering the average inelasticities in
hadron-proton collisions, a carefull analysis of inclusive reaction data was
done by Bellandi {\it et al.} \cite{5}. The average partial inelasticities were
extracted in a model-independent way from experimental data of inclusive
reactions initiated by $ pp$ collisions $(pp\rightarrow c$, $c=\pi ^{\pm
},K^{\pm },p,\overline{p})$. From the results obtained for $\pi ,K,$ $p, 
\overline{p}$ partial inelasticities it was also estimated the behavior of
the total proton inelasticity , wich turned out to be rapidly increasing
with energy in the high energies region.

Bellandi {\it et al.} \cite{6},\cite{7},\cite{8} have already discussed this question in connection
with the behavior of the hadronic cascade and extensive air showers in the
atmosphere, showing that the average proton-air inelasticity is also an
increasing function of the energy. A model dependent analysis of the average
inelasticity in \cite{8} was done by means of the so called Interacting Gluon
model (IGM) \cite{9}, \cite{10}, which includes, besides soft gluonic interactions,
semi-hard QCD interactions responsible for minijet production. This model in
the original version \cite{9} had predicted inelasticity decreasing with energy.
With the addition of the semi-hard component \cite{10} the total inelasticity is
an increasing function of the energy \cite{8}.

For proton-nucleus scattering, at low energy, several models for describing
the leading particle spectrum have been proposed (Interacting Gluon model
and Regge-Mueller formalism) \cite{11}, \cite{12}. Here, we shall work in the Iterative
Leading Particle Model \cite{13}, \cite{14} and use the notation of Frichter, Gaisser and
Stanev \cite{15}. In this model the leading particle spectrum in {\it \ $
p+A\rightarrow N$}(nucleon)$+${\it $X$} collisions is built from sucessive
interacions with $\nu $ interacting protons of the nucleus $A$ and the
behaviour is controlled by a straightforward convolution equation. It should
be mentioned that, strictly speaking, the convolution should be
3-dimensional. Here we only considered the 1-dimension approximation.

\section{Inelasticities}

In this model \cite{15} an iterative scheme was used to write the longitudinal
distributions for multiple scattering of nucleons with wounded nucleons.
After $\nu $ collisions the longitudinal distributions are related by means
of the following 1-dimensional Mellin convolution integral 
\begin{equation}
\label{1}M_\nu ^p(x)=\int_x^1\frac{dy}y[S_{\nu -1}^{+}(y)\beta _{\nu
-1}M_{\nu -1}^p(x/y)+S_{\nu -1}^{-}(y)(1-\beta _{\nu -1})M_{\nu -1}^n(x/y)] 
\end{equation}
for protons and
\begin{equation}
\label{2}M_\nu ^n(x)=\int_x^1\frac{dy}y[S_{\nu -1}^{+}(y)\beta _{\nu
-1}M_{\nu -1}^n(x/y)+S_{\nu -1}^{-}(y)(1-\beta _{\nu -1})M_{\nu -1}^p(x/y)] 
\end{equation}
for neutrons. $M_\nu ^{p,n}(x)$ are the proton and neutron distributions
normalized as 
\begin{equation}
\label{3}\int_0^1dxM_\nu ^{p,n}(x)=n_\nu ^{p,n} 
\end{equation}
with $n_\nu ^p+n_\nu ^n=1.$ The numbers $n_\nu ^N$ express the outgoing
nucleon ($p$ and $n)$ multiplicities for $\nu $ wounded target nucleons. The
superscripts $(\pm $ $)$ describe interactions wich preserve and change the
projectile isospin, respectively, and the parameters $\beta _\nu $
specifying the fraction of isospin preserved reactions. The $S_{\nu -1}^{\pm
}(y)$ define the probability of transation of a nucleon with longitudinal
momentum fraction $x/y$ to a state with longitudinal momentum $x$, after $%
\nu -1$ collisions. For the probability functions $S_\nu ^{\pm }(y)$, we have
for the first collision 
\begin{equation}
\label{4}S_o^{\pm }(y)=\frac{M_1^{p,n}(y)}{\int_o^1dyM_1^{p,n}(y)} 
\end{equation}
with appropriate definitions of $M_o^p$ and $M_o^n$ \cite{15}.

In this model it is assumed that we have different inelasticities upon
subsequent collisions. Adopting a power law form with adjustable factor for $%
\nu >1$, we write
\begin{equation}
\label{5}
S_\nu ^{\pm }(y)=\frac{y^{\alpha _\nu }M_1^{p,n}(y)}
{\int_o^1dyy^{\alpha _\nu }M_1^{p,n}(y)} 
\end{equation}

In order to compute the average nucleon-nucleus elasticity, $<x>_{N-A}$ we
use the Glauber model \cite{glauber}. The $N-A$ leading particle can be obtained by means of the relation 
\begin{equation}
\label{6}
M_{N-A}=\sum_{\nu =1}^\infty P_\nu M_\nu , 
\end{equation}
where $P_\nu $ is the probability of $\nu $-fold collisions of the nucleon
inside the nucleus, given by 
\begin{equation}
\label{7}P_\nu =\frac{\int d^2bP_\nu (b)}{\sigma _{in}^{N-A}} 
\end{equation}
and 
\begin{equation}
\label{8}P_\nu (b)=\frac 1{\nu !}\left[ \sigma _{tot}^{pp}AT(b)\right] ^\nu
\exp [-\sigma _{tot}^{pp}AT(b)] ,
\end{equation}
where $T(b)$ is the nuclear thickness, given in terms of the nuclear density
by
$$
T(b)=\int_{-\infty }^{+\infty }dz\rho (b,z) 
$$
and normalized in the following way
$$
\int d^2bT(b)=1 
$$
The inelastic cross section $\sigma _{in}^{N-air}$ is given by the following
Glauber model \cite{glauber} relation 
\begin{equation}
\label{9}\sigma _{in}^{N-A}=\int d^2b\left[ 1-\exp [-\sigma
_{tot}^{pp}AT(b)]\right] 
\end{equation}

As in \cite{15} we shall assume that the $S_\nu ^{\pm }(y)$ are the same for all
interactions with more than one collision, $\nu > 1$. Using Eqs. (\ref{1}) and (\ref{2}),
it is straigthforward to show that the serie in Eq. (\ref{6}) is an absolutely convergent serie and we can write 
\begin{eqnarray}
(1-<x>_{N-A}) & \equiv & <K>_{N-A} \nonumber \\
&=& \frac {1}{\sigma _{in}^{N-A}} \int d^2b[ 1- \{\eta \exp
[-(1-k)\sigma _{tot}^{pp}AT(b)]+ \nonumber \\
& &  +(1-\eta )\exp [-\sigma_{tot}^{pp}AT(b)]\}]
\label{10}
\end{eqnarray}
where 
\begin{equation}
\label{11}k=\beta _{\nu -1}\int_o^1dyyS_\nu ^{+}(y)+(1-\beta _{\nu
-1})\int_o^1dyyS_\nu ^{-}(y) 
\end{equation}
and 
\begin{equation}
\label{12}\eta =\frac{n_1^p<x>_1^p+n_1^n<x>_1^n}k=\frac{<x>_N}k 
\end{equation}
This parameter $\eta $ defines the relationship between the nucleon
elasticity in the first interaction and the one for the sucessive
interactions of protons and neutrons with nucleus. The Eq. (\ref{10}) gives a
relation between the inelasticity in a nucleon-nucleus collision with
inelasticities in nucleon-nucleon scattering. In a recent paper \cite{16} we have
used this model to analise cosmic ray data on the hadronic flux and we have
shown that the preserved momentum fraction of the sucessive interactions is
the same as the one for first interactions, that is $k\approx <x>_N$, $\eta
\approx 1.$ We then simply have
\begin{eqnarray}
(1-<x >_{N-A})&=&<K>_{N-A} \nonumber \\
&=&\frac 1{\sigma _{in}^{N-A}}\int
d^2b\left[ 1-\exp [-(1-k)\sigma _{tot}^{pp}AT(b)]\right] \label{13}
\end{eqnarray}
In this situation $1-k=<K>_N$ and Eq. (\ref{13}) gives a relationship between
average inelasticities \cite{15}.

It is clear from this relatioship that only in small {\protect $\sigma _{tot}^{pp}$ }
limit is $<K>_{N-A} \quad \simeq <K>_N$. In general, $<K>_{N-A}\geq <K>_N$, the
effect increasing with the increase of $\sigma _{tot}^{pp}$.  If $%
<K>_N \rightarrow 0$, one also has $<K>_{N-A}\rightarrow 0$. On the other
hand, if $<K>_N=1$, then $<K>_{N-A}=1$, and Eq. (\ref{13}) coincides with Eq.
(\ref{9}).

We use here the Woods-Saxon model \cite{17} for the nuclear distribution which is given by 
\begin{equation}
\label{3}\rho (r)=\rho _o\left[ 1+\exp [\frac{r-r_o}{a_o}]\right]
^{-1}(1+\omega \frac{r^2}{r_o^2}) 
\end{equation}
where the factor $(1+\omega \frac{r^2}{r_o^2})$ corresponds to the Fermi
parabolic distribution correction. The parameter $\rho _o$ is a
normalization factor, 
\begin{equation}
\label{4}\int d^3r\rho (r)=1
\end{equation}
The parameters $r_o,a_o$ and $\omega $ can be derived from experimental data
and we have $r_o=0.976A^{1/3}$ fm, $a_o=0.546$ fm and the parameter $\omega $
is given by 
$$
\begin{array}{c}
\omega = -0.25839 \quad if \quad A \leq 40 \\ 
\\
\omega = 0 \quad if \quad A > 40 
\end{array}
$$

In order to calculate $<K>_{N-A}$ we use for $<K>_N$ the values calculated
by means of the Regge-Mueller formalism \cite{12} and as input for $\sigma
_{tot}^{pp}$ we have used the UA4/2 parametrization for the energy
dependence \cite{18}. In the Fig. (1) we show the results of this calculations for
the following nuclei: C, Al, Cu, Ag, Pb and air (A=14.5). In this figure we
also show recent emulsion chamber data for $p$-Pb, $<K>=0.84\pm 016$ \cite{19} and for $p$-C, $<K>=0.65\pm 0.08$ \cite{20}.

In the Fig. (2), we compare the calculated $<K>_{p-air}$ with results from some
models used in Monte Carlo simulation \cite{21}; the Kopeliovich {\it et al.} \cite{22}  (KNP)
QCD multiple Pomeron exchanges model; the Dual Parton model with sea-quark
interaction of Capella {\it et al.} \cite{23}; the statistical model of Fowler {\it et al.} \cite{24} and with calculated values derived from cosmic ray data by Bellandi {\it et
al.} \cite{25}. We note that the calculated $<K>_{p-air}$ in \cite{25} was done
assuming for the $T(b)$ nuclear thickness the Durand and Pi model \cite{26}, which
gives small values for the average inelasticity. In the Fig. (2) we also show
the average inelasticity values as calculated by means of this model.

\section{Conclusions}

We have here calculated the average proton-nucleus ineslaticity in the
Glauber framework, relating the leading particle distributions in
nucleon-nucleus interactions with the respective one in nucleon-proton
collisions. We have compared our results with recent emulsion chamber data
for $p-Pb$ [20] and for $p-C$ \cite{20} at $p_{lab}=1.20\times 10^7$ $GeV/c$. At least in the experimental errors
limit our calculation is in agreement  with these experimental data. We have
also calculated the average $p-air$ inelasticity in a wide range of energy. 
In order to describe the nuclear thickness we have used two models: the
Woods-Saxon model \cite{17} and the Durand and Pi model \cite{26}. The average
proton-air inelasticities calculated by means of the Woods-Saxon model are
larger than the ones calculated by using of the Durand and Pi model. 

One remark should be stressed. The calculated $<K>_{p-air}$values derived from
cosmic ray data \cite{25} were obtained assuming an approximation for the leading
particle distribution in proton-air collisions. Therefore, it is model
dependent. The discrepancies between the values of the  $<K>_{p-air}$ at low 
$\sqrt{s}$ are consequence of the fact that  two different sets
of experimental data were used: nucleonic flux and  hadronic flux at sea level ( for
discussions see \cite{25}). Finally, we note that the calculated $<K>_{p-air}$
with the Woods-Saxon model shows a behavior with $\sqrt{s}$ wich goes
between the values calculated by means of the QCD multiple Pomeron exchanges
model (KNP) and that one calculated by means of the Dual Parton model.

\medskip\ 

We would like to thank the Brazilian governmental agencies CNPq and CAPES
for financial support.

%%\newpage

\newpage

%%\noindent {\large \bf Figure (1) }

\vspace{3cm}

\begin{figure}[htbp]
\begin{center}
{\mbox{\epsfig{figure=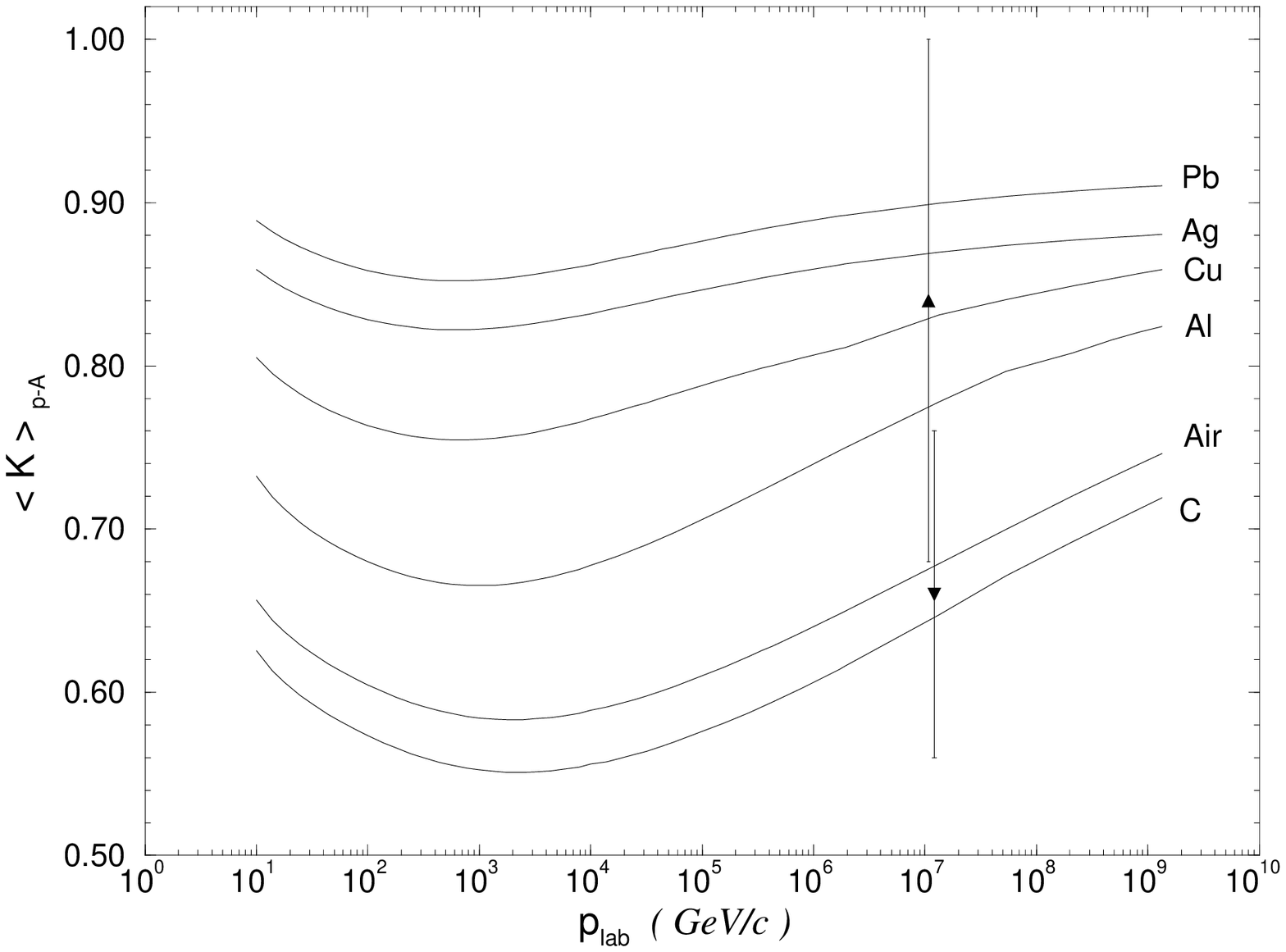,height=10cm}}}
\end{center}
\caption{ Proton-nucleus inelasticities calculated by means of Eq. (\ref{13}). The Pb data (up triangle)  from \cite{19} and C data (down triangle) from \cite{20}.}

\end{figure}

\newpage 

%%\noindent {\large \bf Figure (2) }

\vspace{3cm}

\begin{figure}[htbp]
\begin{center}
{\mbox{\epsfig{file=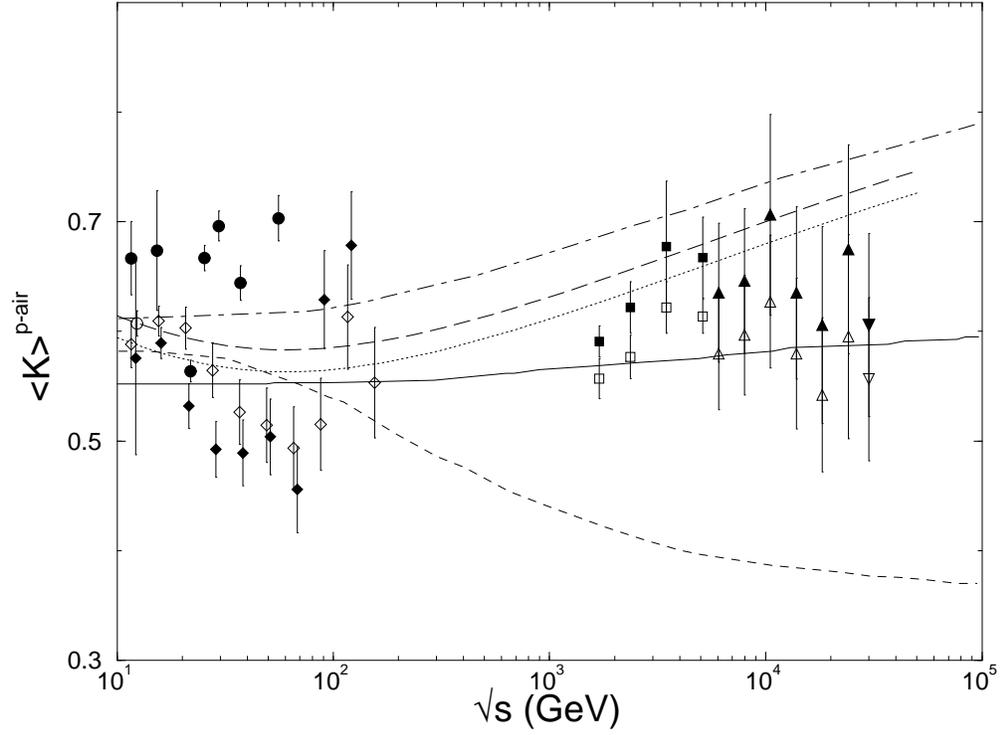,height=10cm}}}
\end{center}
\caption{The $ <K>^{p-air} $ as a function of  $\sqrt{s}$ in $GeV$. The experimental data from \cite{25}. Dashed line from \cite{24}. Full line from \cite{23}. Dot-dashed line from \cite{22}. Dotted line from Eq.(\ref{13}) with Woods
-Saxon model \cite{17}. Long dashed line from Eq. (\ref{13}) with Durand-Pi model \cite{26}.}

\end{figure}

\end{document}